# Spin photocurrent, its spectra dependence, and current-induced spin polarization in an InGaAs/InAlAs two-dimensional electron gas

C. L. Yang[1], H. T. He[1], Lu Ding[1], L. J. Cui[1,2], Y. P. Zeng,[2] J. N. Wang[1], and W. K. Ge[1] *

[1] Department of Physics and Institute of Nano-Science and Technology, The Hong Kong University of Science and Technology, Clear Water Bay, Hong Kong, China

[2] Institute of Semiconductors, Chinese Academy of Sciences, P.O. Box 921, Beijing, China

**Abstract**

Converse effect of spin photocurrent and current induced spin polarization are experimentally demonstrated in the same two-dimensional electron gas system with Rashba spin splitting. Their consistency with the strength of the Rashba coupling as measured from beating of the Shubnikov-de Haas oscillations reveals a unified picture for the spin photocurrent, current-induced spin polarization and spin orbit coupling. In addition, the observed spectral inversion of the spin photocurrent indicates the system with dominating structure inversion asymmetry.

PCAS:    72.25.Fe,    72.25.Pn

With the basic idea of the spin field effect transistor, as proposed by Datta and Das [1], the electric field tunable spin-orbit interaction has been an ideal candidate for manipulating the electron spin-polarization. The structural inversion asymmetry (SIA) induced spin-orbit coupling (Rashba interaction [2]) in a two-dimensional electron gas (2DEG) system has attracted more and more attention due to its potential applications in spintronics devices. With spin splitting of the energy bands, optical excitation of quantum well (QW) structures by circularly polarized radiation will lead to a current whose direction and magnitude depend on the helicity of the incident light. This is named the circular photogalvanic effect (CPGE) and has been demonstrated using inter-subband [3,4] or inter-band [5] excitations. A related effect, in which the current is driven by the spin-flip process of the non-equilibrium population of electrons in the spin-split bands, is called the spin-galvanic effect (SGE) [6]. The most interesting consequence of the $k$-dependent spin splitting is the implication that an applied electric field would induce not only a charge current but also a spin polarization [7-13] along the direction perpendicular to the current.

This letter provides the experimental evidence of circularly polarized optical-excitation-induced spin photocurrent in (001) grown InGaAs/InAlAs 2DEGs under oblique incidence of radiation for inter-band excitation, which is 2 orders of magnitude stronger than similar observations using far-infrared excitation for inter-subband transitions. The current-induced spin-polarization in the same samples is measured using Kerr rotation experiment. These results are consistent with the spin splitting of the energy bands as examined by the beating of the Shubnikov–de Haas (SdH) oscillations. The correlation in these experiments with great consistency reveals a unified picture for the spin photocurrent, current-induced spin polarization and spin orbit coupling. Furthermore, the theoretically predicted spectral inversion of the CPGE spin photocurrent is experimentally observed.

Two samples (named D and E) studied here were $In_xGa_{1-x}As/In_{0.52}Al_{0.48}As$ 2DEGs grown on semi-insulating (001) InP substrate with well thickness of 14 nm. The SIA was achieved by $\delta-$doping of only one side of the barrier layer (on top of the well). To enhance the SIA, sample E was grown with a graded indium composition from 0.53 to 0.75 for the quantum well, instead of the uniform indium composition of 0.70 for sample D. Hall measurements at 1.6 K showed that the carrier concentration was 1.5 x $10^{12}$ $cm^{-2}$ and 1.4 x $10^{12}$ $cm^{-2}$ for samples D and E, respectively. The SdH oscillations as well as their pronounced beating patterns at low magnetic field were indicated in Fig 1 (a) and (b) using the first derivative of $R_{xx}$. The beating pattern arises from two closely separated oscillation frequency components caused by the presence of two kinds of carriers in the system. The beating leads to double-peak structures with similar height in the fast Fourier transform (FFT) spectra and gives the density of the two kinds of carriers as shown in Fig 1(c) and (d). The calculated carrier density is in excellent agreement with the Hall concentration if a spin degeneracy of 1 is taken. In our samples, the distinct beating pattern in the SdH oscillations are believed to come from the spin splitting in 2DEGs, which is further supported by the following spin photocurrent and current-induced spin polarization experiment. The obtained spin-resolved concentrations allow us to determine the Rashba spin-orbit interaction parameter $\alpha$ [14], which gives the value of 3.0 x $10^{-12}$ eVm and 6.3 x $10^{-12}$ eVm for samples D and E, respectively, indicating that the stronger the inversion asymmetry, the larger the spin-orbit interaction in a system. The measured strongly structure-related spin-orbit coupling coefficient suggests that the SIA induced spin-splitting is the dominating mechanism.

The experimental setup for the interband-transition-induced CPGE is schematically shown in the inset of Fig. 2. A tunable Ti-sapphire femtosecond laser was employed for the band-to-band excitation with its polarization modified by either a crystalline $\lambda/4$ wave-plate which yields right ($\sigma^-$) or left ($\sigma^+$) hand circularly polarized light, or a PEM which yields a periodically oscillating polarization between $\sigma^-$ and $\sigma^+$. The electrodes are made at the [110] or [1$\bar{1}$0] sample edges to lead the current along the laboratory *x*-direction. The photocurrent $I_x$ was measured in the unbiased structures at various temperatures via a low-noise current amplifier and a lock-in amplifier. The only difference in our experimental data between those using the $\lambda/4$ wave-plate and those using the PEM is that the former presents a background current even at normal incidence, while the latter does not. This is most likely due to the Damber effect when interband excitation was employed [4]. To show only the polarization dependent signal, all the data presented are measured by PEM, which gives the difference between the $\sigma^-$ and $\sigma^+$ polarizations. The incident light beam is polarized in the *z-y* plane, with an angle of $\Theta$ away from the normal. By changing the angle $\phi$ between the polarization plane of the incident light and the optical axis of the $\lambda/4$ waveplate or PEM, we can measure the helicity dependence of the photocurrent, which is the hallmark of spin photocurrent that distinguishes it from other photocurrent effects.

Figs 2 (a) and (b) show the dependence of the photocurrent of the two samples on the laser beam polarization, represented by angle $\phi$ between the optical axis of the $\lambda/4$ modulator and the polarization plane of the incident light, with λ=880 nm and $\Theta = 30^0$. It clearly shows that when $\phi$ changes from $-\pi/4$ to $+\pi/4$, the current varies from one maximum to the other maximum of opposite direction, in good agreement with the fitting using sin2$\phi$ (see the discussion below). Fig. 2(d) shows the $\Theta$ (incidence angle) dependence of the spin photocurrent. When $\Theta = 0$, since there is no *y*-component of optically induced electron spin polarization, no current is created. With the increase of $\Theta$, leading to a larger *y*-component of the electron spin polarization, the spin photocurrent first gets larger but finally gets smaller because of the increased reflection.

Ganichev et al [4] have systematically used inter-subband excitation to study the CPGE induced spin photocurrent. Inter-band excitation using circularly polarized light will also produce a spin photocurrent, for which the microscopic mechanism can be schematically described as shown in fig. 2(c). Considering the SIA induced Rashba interaction, there will be non-parabolic terms in the Hamiltonian, such as the linear (in $k_{//}$) spin splitting for electrons (or light holes), and spin splitting of the heavy-hole [15] states proportional to $k_{//}^3$. The coupling between the spin and wave vector of the carriers, as well as the spin related selection rules, yield a non-uniform distribution of carriers in k-space upon circularly polarized optical excitation. The imbalanced momentum relaxation of electrons in the conduction band results in a net current due to the absorption of the circularly polarized light.

For quantum well structures of symmetry $C_{2v}$ grown along the principal axis [001] with structural inversion asymmetry, a photocurrent can be generated only under oblique incidence of irradiation. If the incidence is in the (y, z) plane, then the photocurrent is induced along the x direction [3,4] and can be phenomenologically estimated to be $j_x \propto \gamma_{xy} P_{circ} \sin\theta$, where $\gamma_{xy}$ is a pseudotensor component related to the Rashba coupling coefficient, $\theta$ is the angle of refraction ($\sin\theta = \sin\Theta/\sqrt{\varepsilon^*}$, $\varepsilon^* \sim 13$). Since $P_{circ} = \dfrac{I_{\sigma^-} - I_{\sigma^+}}{I_{\sigma^-} + I_{\sigma^+}} = \sin 2\phi$, $j_x$ is expected to be proportional to $\sin 2\phi$ at a fixed incidence angle $\Theta$, just as we have observed. The solid line in Fig. 2 (d) is the fitted result for the $\Theta$ dependence of $j_x$ considering of reflectance, which is in excellent agreement with the experimental data. Furthermore, we find that the photocurrent is almost linearly dependent on the laser power. For sample D, the current drops faster when temperature is higher than 200 K, leaving a very small signal at 300 K. But for sample E, the current only drops about a factor of 4 from 11 K to room temperature. We believe that the larger photocurrent of sample E is due to its much larger spin splitting, as also revealed by the SdH experiment. We also find that the inter-band excitation induced spin photocurrent is up to 2 orders of magnitude stronger than that of the inter-subband transitions in similar experiments

In addition to the incident direction and the polarization dependence of the spin photocurrent, we have found that it also changes its direction when the laser wavelength is changed, as indicated in Fig. 3 for the spectra dependence of CPGE photocurrent. The photo-reflectance spectra of the samples are shown to clearly indicate the quantized energy levels of electrons and holes as marked by the arrows. For the two samples, we have explicitly shown that the CPGE spin photocurrent changes its sign when the laser wavelength changes, which is in agreement with the theoretical prediction by Golub [16]. The spectral inversion in the CPGE spin photocurrent is thought to be a characteristic to distinguish the spin splitting mechanisms between the SIA and the bulk inversion asymmetry (BIA). The big dip in the spectra response suggests that the SIA is the dominating mechanisms for the spin splitting in our samples. This is also consistent with the dramatic change of the Rashba coefficient with our modification of the sample structure as mentioned in the SdH measurement. There are mainly two contributions for spectral inversion of the spin photocurrent. The first one comes from the transitions from the light hole to the electron subband. Due to the selection rules for light hole and heavy hole (such as -3/2 → -1/2 for heavy hole, -1/2 → +1/2 for light hole with $\Delta j = +1$), electrons excited from the heavy hole and the light hole bands by the same circularly polarized light will jump to opposite spin splitting branches (+1/2 or -1/2) in the conduction band, which will result in reversed current in the SIA case. Secondly, as pointed out by Golub [16], the negative contribution can also come from the heavy hole related transitions at larger wave vector due to the mixing or anti-crossing of the hole bands. The spectra inversion in the CPGE spin photocurrent is very simple and useful in distinguishing the spin splitting mechanisms between the SIA and BIA. We intend to carry out further studies along this direction, to involve ground states excitations by changing the sample composition and/or expanding the laser wavelength range, and to make careful calculations for comparison.

There have been theoretical predictions for spatially homogeneous spin polarization resulting from an electrical current [8] in systems such as 2DEGs. Indeed we can simply interpret this current-induced spin polarization as the converse process of the spin photo-current. Since the Rashba coupling term in the Hamiltonian can be expressed as $H_{SO} = \alpha \vec{\sigma} \cdot (\vec{z} \times \vec{k}_{//}) = \alpha \vec{k}_{//} \cdot (\vec{\sigma} \times \vec{z})$, one can interpret this coupling as if the momentum (carried by a current) induces an effective magnetic field, or the spin induces an effective momentum. Inoue $et\ a\ l$ [10] first derived the diffusive conductance tensor for a disordered 2DEG with spin-orbit interaction and showed that the applied bias ($E_x$) induces a spin accumulation $\langle S_y \rangle = 4\pi e \tau D \lambda E_x$, where $D = m_e / 2\pi \hbar^2$ is the 2DEG density of states per spin, the lifetime $\tau$ is the momentum relaxation time, and $\lambda = \alpha \langle E_z \rangle / \hbar$ represents the Rashba interaction. It indicates that the intensity of the current-induced spin polarization will also give a measure of the spin-orbit interaction parameter $\alpha$.

Kerr effect was used to monitor the spin polarization along the *y*-direction under an external electric field in the *x*-direction, as shown in the inset of Fig. 4. In the Kerr rotation experiment, a polarizer and an analyzer were placed in the incident and reflected laser beams, respectively. A PEM situated before the analyzer on the reflected beam was employed to sensitively monitor any optical polarization rotation induced by the spin polarization in the samples. To get rid of any possible reflectivity change induced by heating effects due to the applied electric field, the Kerr rotation was extracted from the signal difference between positive and negative electric fields. The laser wavelength was tuned to be close to the transition energy between the third excited state of heavy hole and the electron, which was identified from photo-reflectance measurements.

Figs. 4(a) and (b) show the Kerr rotation against the external electric field at 11 K for samples D and E, respectively. The detected Kerr signal increases with increase of the external field, which is in agreement with the theoretical prediction that the spin polarization $S_y$ is proportional to $E_x$. However, we didn't observe linear relationship between the spin polarization and external field, which can originate from two possibilities. Firstly, nonlinearity at low field mostly comes from the experimental uncertainty due to the limited sensitivity. The second reason is the electric field dependent spin relaxation time. The spin relaxation time is found to decrease [12] with the increase of electric field, resulting in a reduced spin polarization or spin density, which is the reason that Kerr rotation signal gets saturated at high field. Also we find that sample E gives a Kerr rotation about 3 to 4 times larger than that of sample D, which is consistent with the much larger spin-orbit coupling coefficient in the former as determined by the SdH oscillation. The temperature dependence of the Kerr rotation, as presented in the inset of Fig 4(a) for sample D, shows that the spin polarization drops quickly when temperature is above 100 K, which is due to the reduced spin and momentum relaxation time at elevated temperatures.

In summary, we have shown two converse effects, spin photocurrent and current-induced spin polarization, in a specially designed Rashba system, i.e. a 2DEG system with asymmetric potential barriers. It is worth noticing the comparison of the two samples used in this work. The Rashba coefficient measured from SdH oscillations has a ratio of about 2 between samples E and D. For the spin photocurrent, the efficiency ratio is a bit less than 3 between these two samples. For the Kerr rotation, the relative efficiency ratio is about 3 to 4. This evidently shows that the physical mechanisms behind the three effects are uniformly based on the Rashba coupling, and any of them can be taken as an indication of the Rashba effect. Though the general trends are consistent, we do not expect to find precise values for these ratios, as so many parameters are involved in any quantitative comparison. Finally, the spectral inversion in the CPGE spin photocurrent provides us an easy way to distinguish the SIA and BIA in the spin-orbit coupling.

Acknowledgement: We are grateful to the encouraging discussions with Prof. F. C. Zhang and Prof. S. Q. Shen of the University of Hong Kong. We also greatly appreciate Prof. B. A. Foreman for his help and his critical reading of the manuscript.

# References

* Corresponding Author, email: phweikun@ust.hk

1. S. Datta and B. Das, *Appl.Phys.Lett*., **56**, 665 (1990).

2. Y. A. Bychkov and E. I. Rashba, *J. Phys. C* **17**, 6039 (1984).

3. S. D. Ganichev, E.L. Ivchenko, S.N. Danilov, J. Eroms, W. Wegscheider, D. Weiss, W. Prettl, *Phys. Rev. Lett.* **86**, 4358 (2001).

4. S. D Ganichev and W. Prettl, *J. Phys. C*, **15**, R935 (2003), and references therein.

5. V. V. Bel'kov, S. D. Ganichev, Petra Schneider, C. Back, M. Oestreich, J. Rudolph, D. Hägele, L. E. Golub, W. Wegscheider, W. Prettl, *Solid State Commun*. **128**, 283 (2003).

6. S. D. Ganichev, E. L. Ivchenko, V. V. Bel'kov, S. A. Tarasenko, M. Sollinger, D. Weiss, W. Wegscheider and W. Prettl, *Nature*, **417**, 153 (2002).

7. A. G. Aronov, Y. B. Lyanda-Geller, and G. E. Pikus, *Sov. Phys*. JETP **73**, 537 (1991).

8. V. M. Edelstein, *Solid State Commun*. **73**, 233 (1990).

9. V. Chaplik, M.V. Entin, and L. I. Magarill, *Physica* E **13**, 744 (2002).

10. J. I. Inoue, G. E.W. Bauer, and L.W. Molenkamp, *Phys. Rev*. B **67**, 033104 (2003).

11. Y. K. Kato, R. C. Myers, A. C. Gossard, and D. D. Awschalom, *Phys. Rev. Lett*. **93**, 176601 (2004).

12. A. Yu. Silov, P. A. Blajnov, J. H. Wolter, R. Hey, K. H. Ploog, N. S. Averkiev, *Appl. Phys. Lett*., **85**, 5929 (2004).

13. S.D.Ganichev, S.N.Danilov, Petra, Schneider, V.V.Bel'kov, L.E.Golub, W.Wegscheider, D.Weiss, and W.Prettl, *cond-mat*/0403641 (2004).

14. G. Engels, J. Lange, Th. Schapers, and H. Luth, *Phys. Rev*. B **55**, 1958 (1997); *J. Appl. Phys*. **83**, 4324 (1998).

15. R. Winkler, *Phys. Rev*. B **62**, 4245 (2000).

16. L. E. Golub, *Phys. Rev*. B **67**, 235320 (2003).

**Figure Captions:**

Fig. 1 (a) and (b) are the beating patterns of the SdH oscillations at 1.6 K under low magnetic field as shown in the first derivative of the magneto-resistance $R_{xx}$ of samples D and E, respectively. (c) and (d) are the calculated carrier concentrations based on the FFT spectrum from (a) and (b).

Fig. 2 (a) and (b) are the angle $\phi$ dependence of the spin photocurrent in the 2DEG sample D and E, respectively, with oblique incidence angle of $30°$ and laser power of 100 mW ($\lambda = 880$ nm) at 10 K, showing the helicity dependence of the photocurrent. The solid lines are the fit using $\sin 2\phi$. (c) is a schematic diagram with spin-split valence and conduction bands to show the microscopic origin of the CPGE induced photocurrent. The vertical arrows show the permitted transitions with $\sigma^-$ excitation and the horizontal arrow indicates the final direction of the induced photocurrent. (d) is the incidence angle $\Theta$ dependence of the photocurrent. The solid line is the fitted curve. The experimental setup of the spin photocurrent is depicted in the inset.

Fig. 3 Spectral response of the spin photocurrent for (a) sample D and (b) sample E. The photo-reflectance spectra of the two samples are also shown to determine the electronic structures of the samples. The arrows indicate the heavy hole (hh) and light hole (lh) related transitions.

Fig. 4. Electric field ($E_x$) dependence of the Kerr rotation of sample D (a) and E (b) to show the current induced spin polarization $\langle S_y \rangle$ at 10 K. The temperature dependence of the Kerr rotation of sample D ($E_x = 50$ V/cm) is shown in the inset of Fig. 4(a). The experimental setup is schematically shown in the inset of Fig. 4(b), of which P1 and P2 represent the polarizer and analyzer, respectively.

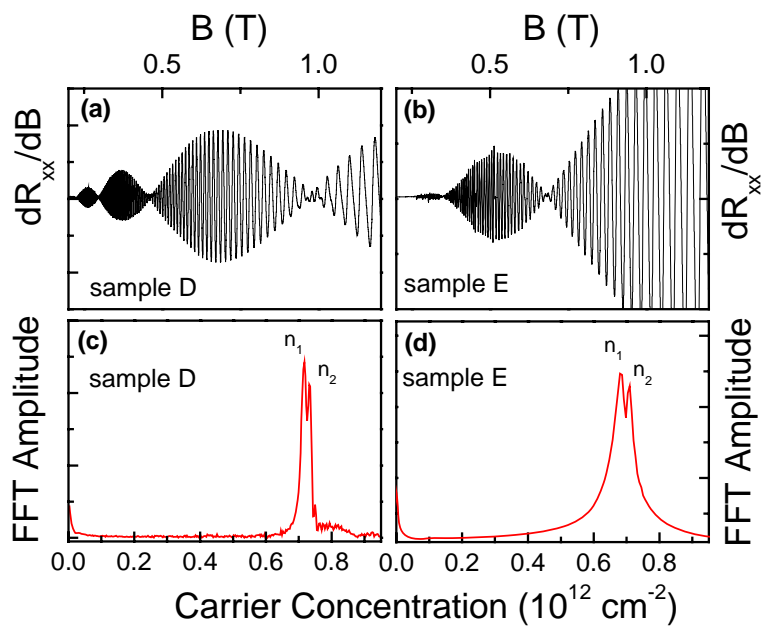

Figure 1

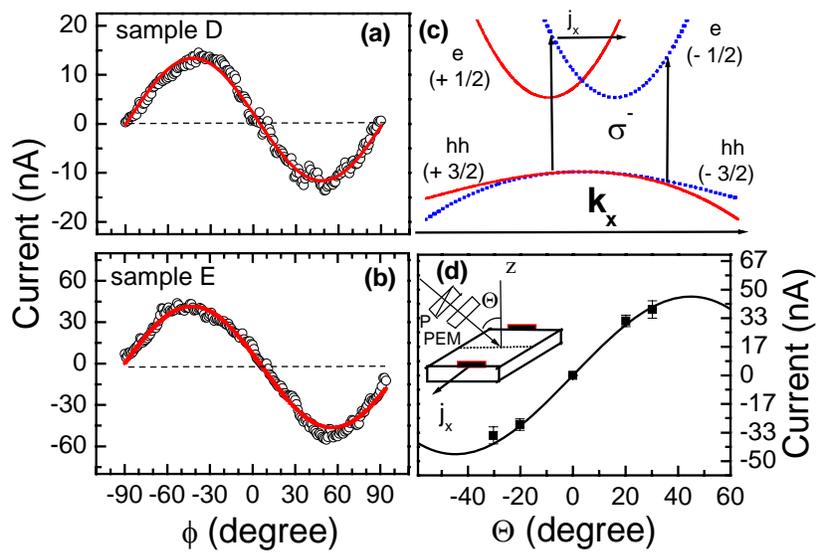

Figure 2

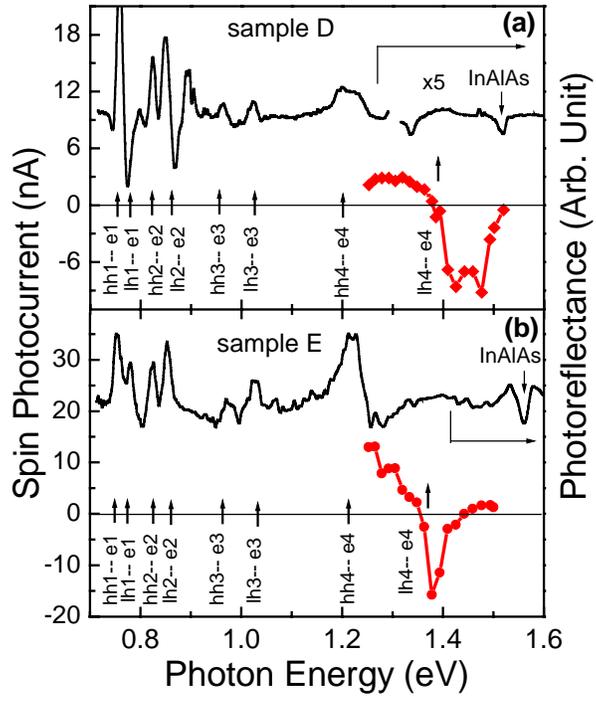

Figure 3

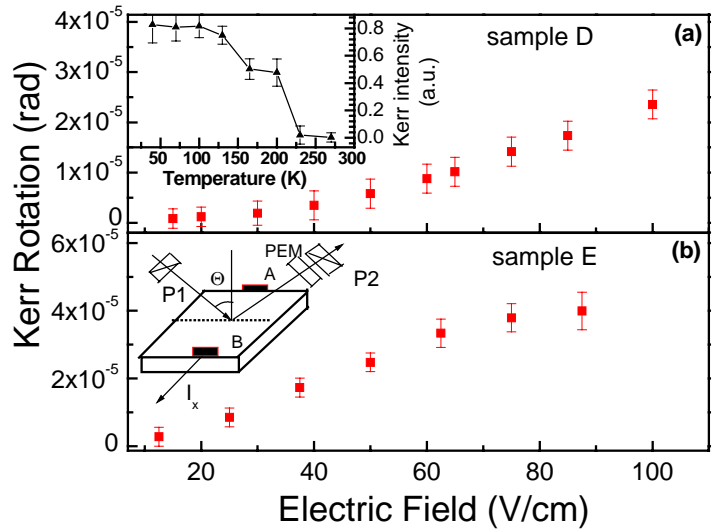

Figure 4

13